# Semantic Matching of Security Policies to Support Security Experts

Othman Benammar, Hicham Elasri, Abderrahim Sekkaki
University Hassan II, Faculty of Sciences Ain Chock
Casablanca, Morocco
emails{o.benammar@live.fr, h.elasri@ancfcc.gov.ma, a.sekkaki@fsac.ac.ma}

*Abstract*— **Management of security policies has become increasingly difficult given the number of domains to manage, taken into consideration their extent and their complexity. Security experts has to deal with a variety of frameworks and specification languages used in different domains that may belong to any Cloud Computing or Distributed Systems. This wealth of frameworks and languages make the management task and the interpretation of the security policies so difficult. Each approach provides its own conflict management method or tool, the security expert will be forced to manage all these tools, which makes the field maintenance and time consuming expensive. In order to hide this complexity and to facilitate some security expert's tasks and automate the others, we propose a security policies aligning based on ontologies process; this process enables to detect and resolve security policies conflicts and to support security experts in managing tasks.**

## I. INTRODUCTION

The security management in Cloud Computing (CC), Grid Computing (GC) or Distributed Systems (DS) requires the use of a dynamic environment-based policy to overcome a set of problems like the number of domains to manage and the permanent modifications that could occur inside managed environments themselves. However, the main aims for the management-based policy are to optimize, minimize the human interventions and automate some management tasks during security lifecycle.

Several frameworks of security-based policy and models have been proposed in the literature. The present paper will focus on the languages associated to KAOS [10], REI [11] and Ponder [5].

The main issue to manage many heterogeneous security domains is to assure interoperability between the enforced Security Policies (SP), and also detect and resolve conflicts.

Conflicts can be categorized in two forms, vertical and horizontal conflicts. The vertical ones occur when the SP are expressed in a common language while sharing heterogeneous knowledge on shared resources, this type of conflict is typically semantic. The horizontal conflicts occur when security policies are expressed with different languages; this type of conflict is due to different syntax for used languages. However, to detect and resolve horizontal conflicts, a third-party canonical language can be used to uniformize and hide syntactic and lexical heterogeneity; as a result, we find the vertical form.

Ontologies are the most used approach to represent semantic knowledge, and thereby provide robust mechanisms to resolve semantic conflict. In this sense, we use the ontologies as a pivot between languages towards to facilitate interpretation and understanding for the security experts.

Our goal is to provide a helpful method and tool to hide the complexity to deal with multiple specification languages that security expert can find to manage several security domains like those present in DS, GC or CC. So, we define an ontology used as a common knowledge, called in the literature support or background ontology [3]. This ontology is a semantic consensus on the concepts of different security policy specification languages. It is used in a distributed environment to supply security domains with a mutual agreement on the SP which implies a secured semantic interoperability. This paper presents an aligning-based ontologies process that supports security experts to manage SP , assures the SP conformity and consistency, detects and resolves conflicts that may be occur when adding, changing or merging SP.

This work will be presented as follows. The second section presents some related researches. The third section explains briefly the management-based policy principle. The fourth section will display the most used security-based policy frameworks such as KAOS, REI and PONDER. The fifth section presents our SP alignment process approach by using examples for illustration and a prototype is given in the sixth section. Finally, the conclusion will be featured in the last section.

## II. RELATED WORK

In this section, we briefly present some of the research literature related to some security-based policy frameworks.

On one hand, there are some works in detecting conflicts of security policy related to syntax conflicts. Ponder and XACML [27] are typically non-semantic policy framework. On the other hand, there are some representative research works on semantic conflict detection. KAOS and REI are approaches that enriched with semantics using RDF [28] and OWL [25] as standards for policy specification. KAOS adopt policy priority for the conflict resolution of semantic policy [29]. REI combined the method of meta-policy and priority for conflict detection and resolution of semantic policy [18]. A comparative analysis between semantic and non-semantic language is made by Nejdl et al. [30] to show the advantages of semantic policy approach.

Since each semantic approach provides a conflict management method, the security expert will be forced to manage all these tools. Some works was interested to detecting conflict between SP related to a specified framework [18][23]. Our goal is to interface with different frameworks in order to centralize the conflicts management and offer the possibility to compare SP from different frameworks in order to detect and resolve conflicts. This work cover the first step that present an ontology-based method in order to remove all semantic SP ambiguity, assist security experts to resolve semantic conflicts and enable an automatic resolution for semantic conflicts by adopting a conflict resolution based on rules strategy i.e., by combining a resolution rules set for each conflict type.

### III. MANAGEMENT-BASED POLICY PRINCIPLE

The objective of the management-based policy is the optimization as well as possible of security experts efforts. Thus, it first consists in determining the strategies and the tactics reflecting the security expert's objective and also representing them in policies form. Then, these policies must be presented as a set of rules to be understood by the management entities and stored in a Policy Repository (PR). The distribution and the application of these policies require to communicate these rules to the PDP (Policy Decision Point) and the PEPs (Policy Enforcement Point) managed by this latter. The high-level policy defined by security expert, cannot be directly understood by equipments or applications. It is therefore necessary to translate it into the configuration rules. At each stage, several policies may conflict. There is no simple recipe for solving all conflicts encountered. Ultimately, the security expert must perform this task.

The term subject refers to users, principals, or automated manager components, which have management responsibility (i.e., have the authority to initiate a management decision). The term target refers to objects on which the action can be performed. The relations between subject and target [12] are well defined by management policies and depend also on the nature of these latter. Thus, obligation policies define what a Subject must perform or not on the level of a Target, whereas authorization policies specify the access rights that could have a Subject on the level of a Target.

The policies rules database is replaced by both a domain service and a policy service. In our approach, the policies specification will be based on the ontologies, to ensure a horizontal and vertical semantics between SP specification languages.

In the next section, we present the security policies semantic specification languages chosen through literature.

### IV. POLICY SPECIFICATION LANGUAGES AND FRAMEWORKS

SP can be specified in several ways and different approaches have been proposed in various application fields [16]. However, there are some general requirements that any policy representation must satisfy independently of its application scope:
– A great expressiveness that can meets the managed system security requirements
– A simple and effective vocabulary to facilitate to the security experts policy development tasks with various expertise degrees.
– An execution space that provide a political specifications mapping in different platforms.
– Scalability to ensure appropriate performance, enable the defined policies analysis and achieve an appropriate balance between the expression objectives, its calculating docility and ease of use.

This section does not provide security policies specification overview, but to describe the semantic aspects of some approaches that have been specifically designed and tested for the distributed systems management. We first present some semantic security approaches KAOS, REI and Ponder.

#### A. KAOS (Knowledge-able Agent-oriented System)

KAOS is a framework that provides the policy services allowing the specification, management, conflicts resolution, and execution of policies within fields. Policies are represented in DAML + OIL [2] as ontology. Policy ontologies KAOS distinguish between authorizations (i.e., constraints that permit or prohibit certain actions) and obligations (i.e., constraints that require an action to perform or to waive such requirement).

KAOS detects potential conflicts between the policies at the specification moment, every time a user tries to add a new policy to the directory service [15]. The engine identifies conflicts between policies using the subsumption mechanisms between classes and tries to resolve these conflicts through political order based on their priority and, if necessary, create new harmonized policies [17][18].

#### B. REI

REI is a framework that includes specifying policies support, analysis and reasoning in computer applications [11][12].

His language policy based on ontological logic allows users to express and represent the rights concepts, prohibitions, obligations and exemptions. These concepts correspond, respectively, to the positive and negative authorization conditions, and to positive and negative obligations in Ponder and KAOS. REI is based on an application-independent ontology to represent the rights concepts, prohibitions, requirements, exemptions and the politics rules. This allows different components from the same domain to understand and interpret REI policies in the right way. REI's policy specification can be in forms of Prolog predicates or RDF-S [31] statements. REI extends OWL with the expression of relations like role-value maps, which makes the language itself more expressive than original OWL.

## C. Ponder

Ponder is a declarative object-oriented language that supports several types of distributed systems management policies specification. It's providing technical structuring policies to meet the policy administration complexity in large company information systems [5][8][9]. The basic types of political rights in Ponder are obligations and permissions.

Ponder has a conflict detection tool to detect overlaps and conflicts between policies. Similar to KAOS and REI, the tool can detect inconsistencies in the policy specification that may arise between policies with opposite signs modalities which refer to the same subjects, targets or actions (e.g., conflict between permissions [4]).

Following is an SP comparative table for semantic specification languages based on several criteria.

TABLE I. SPECIFICATION LANGUAGES COMPARISON OF SEMANTIC SECURITY POLICY

|   | KAOS | REI | Ponder | Our approach |
|---|---|---|---|---|
| **ontology-based** | yes | yes | no | yes |
| **policy-type** | Access control based | Access control based | Access control based | Access control based and management |
| **Policy representation** | Owl | Rei (Prolog like syntax+ RDF-S | Ponder language specification | OWL |
| **Open security interoperability support** | Applying mediating and proxy agents on to specific domains | Applicable in specific domains | Applicable in specific domains | Applicable in specific domains |
| **Interoperability representation** | Explicit | Explicit | Explicit | Explicit |
| **Reasoning support** | Java theorem | Prolog engine | Event calculus representation | Java theorem |
| **Security ontology enrichment** | no | no | no | yes |

Our goal is to ensure unified presentation of different semantics languages and other SP specifications through a specification languages transformation set to a single ontological representation. This transformation can resolve vertical level conflicts by hiding the heterogeneous knowledge semantic heterogeneity and horizontal level conflicts by hiding the languages heterogeneity. Since each approach KAOS, REI, or Ponder provides a conflict management method, the security expert will be forced to manage all these tools from different approaches. In order to hide this complexity, we also manage the horizontal level conflicts between different languages to facilitate some security expert tasks and automate the others. In the present work we focus on the semantic conflicts resolution tasks between SP. We will present in this work an ontology-based method in order to remove all semantic SP ambiguity, assist security experts to resolve semantic conflicts and enable an automatic resolution for semantic conflicts by adopting a conflict resolution based on rules strategy: by combining a resolution rules set for each conflict type.

### D. Types of conflicts

To harmonize the SP, we must lift any type of conflict that we enumerate into three categories: syntactic, structural and semantic conflicts. The first conflict type concerns identity or abbreviation conflicts introduced during SP designing or primitives like operations and attributes. The second one involves structural conflicts between classes. Here we distinguish between those related to association type between two elements and those related to the hierarchy relative to an abstraction level. The third conflict type concerns the conflicts that involve semantic elements like naming conflicts, measurement and confusion. These conflict types can occur in two forms vertical and horizontal form.

– Vertical conflicts occur when we have SP expressed in a common language while sharing heterogeneous knowledge on shared resources, this type of conflict is typically semantic.
– Horizontal conflicts, they occur when security policies are expressed with different languages, this syntactic type of conflict is due to different syntax languages.

## V. SEMANTIC ALIGNMENT PROCESS OF SECURITY POLICIES

In our study, we could see that the ontology languages facilitate understanding between interoperate entities. However, interoperability access controls cannot be implemented without the semantic compatibility establishment between the SP agents participating in the exchange. One of the problems is heterogeneity between different organizations entities, this heterogeneity makes the semantic compatibility establishing so difficult. The scientific community became interested in ontological models for the access control purposes. The most significant models in our view are KAOS [10] and REI [13][20].

In this section, we present our SP alignment process that can possibly be described with semantic specification languages like KAOS, REI or others.

### A. Security policy semantic alignment process

The semantic alignment process allows detecting and resolving naming semantic conflicts type between candidate SP for cooperation in distributed systems, and it enable to enhance the security ontology used as support ontology during the alignment process.

The alignment process inputs are:
- A SP set, denoted $SP_1$, $SP_2$ ... $SP_n$, selected by the security expert for their alignment.
- A security ontology chosen by the security expert according to SP type used in the distributed system. It will serve as support ontology during the alignment process.

The alignment process provides as outputs:
- A new enriched security ontology by new semantic relations added due to two treatments that we apply in the process: ontologies alignment and enrichment, which can be used to update the initial security ontology and serve as new support ontology in future integration iterations thereby increasing the process efficiency.
- Correspondence ontology, security experts can use this ontology to detect and resolve semantic conflicts in a semi-automatic process.

The semantic alignment process is highlighted in Figure 1.

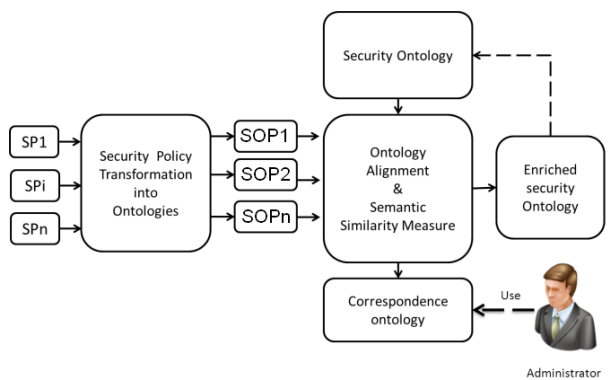

Figure 1. Matching security policy process.

Our proposal is based on several research projects results, especially those concerning the SP transformation described in the SP specification languages in ontologies description languages [7], and those related to the ontologies security alignment and enhancement [3][7][19][20]. The security policy transformation into ontologies relies on the work cited above to transform SP to Security Ontology Policies (SOP).

The integration process involves two steps:

1. The candidate security policies for alignment-based ontology

The candidates SP for alignment are conceptual safety rules set representations. They are described in a given modelling language (formal, semi-formal or natural).

Several studies have focused on the conceptual models transformation [7][21][22][24]. We rely on this work for the SP transformation into ontologies.

2. Aligning obtained ontologies based on security ontology

The ontologies alignment consists on establishing correspondences between two ontologies, with a priori on the same knowledge domain. It helps to find semantic relations between concepts defined in ontologies to align.

In order to achieve semantic SP alignment, we rely on an ontology aligning method [1] based on support ontology, also called in the literature "background ontology" [3].

This step takes as input:
- Ontologies obtained in the previous step.
- The security ontology.

Results:
- An ontology alignment results of all available ontologies as input.
- The ontology of security enhanced by new relationships.

Support Ontology is in our case, the SP ontology used by the distributed system.

### B. Security ontology enrichment

The enrichment allows identifying new elements: concepts, terms and relations, then place them in an existing ontology. Enrichment and manual ontology construction prove to be tedious and expensive tasks [3]. Several studies have proposed automated methods or semi-automated enrichment and ontologies construction. Most of these methods rely on external sources from which new semantic knowledge are identified, evaluated and placed in the ontology to enrich it.

The method for measuring semantic similarity [1] uses security ontology enrichment process, when it has no semantic relation between concepts to align. To implement this process and demonstrate its feasibility, we chose, and that as an example, two rules among the various rules on semantic relationships:

- R1: Two concepts are similar if their neighbours are similar equivalents. According to [24], two concepts are similar if their sub-concepts "son" are similar. This was confirmed in [6].
- R2: Two concepts are similar if their sub-concepts "son" are similar.

Rule R2 focuses on the composite concepts. The composite concepts represent concepts Father and sub-components concepts linked by a semantic relationship like "part-of", are the concepts son.

Let C1 and C2 two concepts to align belonging to $SOP_i$; we distinguish three cases:

Case 1: C1 and C2 recognize a semantic relationship within $SOP_i$. This relationship is then injected into the Security Ontology (SO).

Case 2: C1 and C2 do not allow $SOP_i$ in semantic relation while there are two concepts in $SOP_i$, C1' and C2' and two semantic relations of equivalence, the first between C1 and C1' and the second between C2 and C2'.

From R1 we can deduce a new semantic relationship between C1 and C2 that is injected into the security ontology SO.

Case 3: C1 and C2 are concepts that do not allow composite semantic relation in $SOP_i$, while there are semantic relations between concepts in their respective son.

Let {C11, C12 ...C1n} the C1 son concepts set and {C21, C22 ... C2n} the C2 son concepts set, as is C1i and C2i admit a semantic relationship within $SOP_i$. From R2, we can deduce a new semantic relationship between C1 and C2 that is injected into the security ontology SO.

### C. Semantic and syntactic similarity measuring

To achieve the ontologies alignment for the various SP, we reuse methods for measuring semantic and syntactic similarity proposed in [1], which proposes several methods for measuring semantic and syntactic similarity.

Ontologies alignment process product aligned ontology, which defines the semantic relationships between several sources ontologies concepts. Obviously, we need a model that combines semantic relationship to each one or more rules, they can be different types: transformation rules, merging, mapping, renaming or deleting...

### D. Conflict resolution rules

To demonstrate how to use correspondence ontology, we present resolution rules for naming conflicts derived from semantic relation: homonym and synonym existing in correspondence ontology result of our process.

**Conflict Resolution Rule 1:** if we have a semantic relation type synonym in the correspondence ontology between concepts of sources ontologies, we offer security experts to rename the concepts with same name.

**Conflict Resolution Rule 2:** if we have a semantic relationship type homonym in the correspondence ontology between concepts of sources ontologies, we offer security experts to rename the concepts with different names.

Figure 2 shows a namely conflict resolution assisted by security experts based on a set of conflict resolution rules stored in a catalogue.

Based on correspondence ontology and the conflict resolution rules, we offer security expert decisions set represented by derived operations set. For example, in case of type synonym relationship in correspondence ontology then find in the catalogue the resolving conflicts (conflict resolution rule 1), then propose to security expert an operation "rename" one of concepts in conflicts.

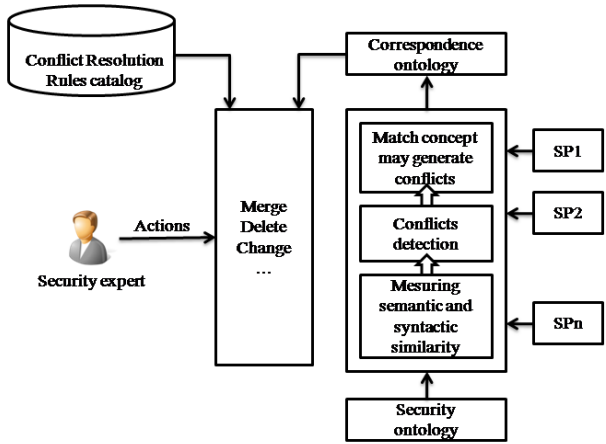

Figure 2. Matching security policy process with actions.

### E. Matching security policies in Cloud Computing environments

Cloud computing providers can build large datacenters at low cost due to their expertise in organizing and provisioning computational resources. The economies of scale increase revenue for cloud providers and lower costs for cloud users.

Several entities intend to take part in the benefits of Cloud Computing by outsourcing their infrastructure while keeping their policy and operational security system. To illustrate our process of matching the SP, we present the case of several entities wishing to outsource their infrastructure to a Cloud Computing and using different security policies expressed in different SP specification languages. Our approach offers to the Cloud Computing security experts a tool to manage these different SP while ensuring consistency with the Cloud Computing security policy if necessary. For this, we propose to align the different SP to provide a unified global vision and thus detect potential conflicts between these SP. The security expert can rely on an ontology correspondence as a decision support to detect synonyms and homonyms concepts; this example is highlighted in Figure 3. This process is essential toward a centralized management tool in CC environments.

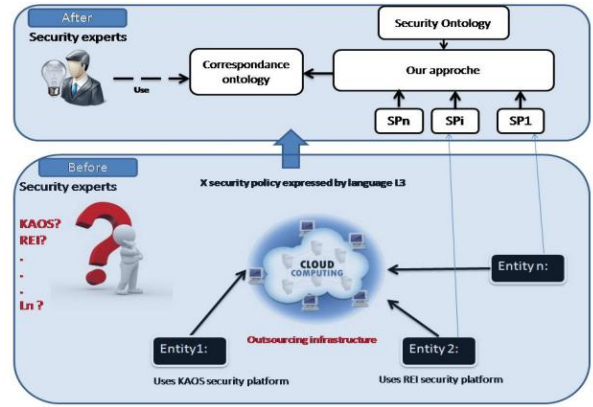

Figure 3. Matching security policy example

In the follow section, we show an example to illustrate the ontologies alignment process presented by Figure 2, and validate the function of this process. For example, as shown in Figure 4, we suppose that the collaborative environment is composed by two virtual domains. So, the input contains two cases which are obtained from domain A and domain B, each case describes the positive authorization policy in REI Prolog format, which specifies that an entity X (P in case 1 and Q in case 2) can use the printing service when it is an IT department member. The part of the instances describing two cases is as follow:

**Case 1 from domain A**

The following is a positive example for authorization policy in REI Prolog format [32].
has(P,permit(usePrintingService,[member(P, ITDepartment)])).

The example above specifies that an entity P can use the printing service when it is an IT department member in domain A.

**Case 2 from domain B**

We consider the same proposals described in case 1 by replacing the entity P by Q and domain A by domain B.

has(Q, allow(usePrintingService,[member (Q, ITDepartment)]).

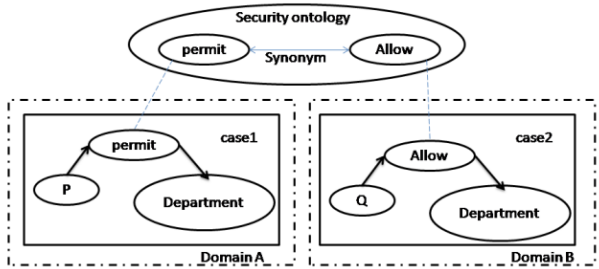

Figure 4. The two cases represented fragments ontologies

We apply the matching process of aligning two cases 1 and 2 from domain A and B, we obtain a correspondence ontology that represents the tow concepts in conflict(synonym(permit, allow)) and based on correspondence ontology to derive the rule of conflict resolution to use. In this example it is the conflict resolution rule 1 then proposes to security experts an operation "rename" one of the component concepts by the same concepts.

The example, shown in Figure 4, allows presenting how to solve the naming conflict between the concepts from the two cases specified by the security policy: REI. Our approach can solve all conflicts types: horizontal and vertical knowing that we define the resolution rules for each relationship between concepts in the correspondence ontology.

## VI. PROTOTYPE

In order to validate and to evaluate our process of semantic matching of security policies, we have developed a prototype baptized Semantic Matching of Security Policies (SMSP). The prototype SMSP takes first in entry the domain security ontology and the security policies to match; and transforms them into ontologies described in OWL [26].

Security expert chooses the appropriate actions to perform through a list of rules already established in order to resolve the semantic conflicts. Then, SMSP applies various treatments associated with each step of our proposed method, in particular similarity measurement and security ontology enrichment. After those treatments, domain security ontology were be enriched with new semantic relationships toward to be used in similar situations.

Figure 5 presents a screen capture of SMSP that shows a use case for matching security policies; user can load domain ontology and the two security policies to match, and perform an action by merging, renaming or deleting conflicting concepts.

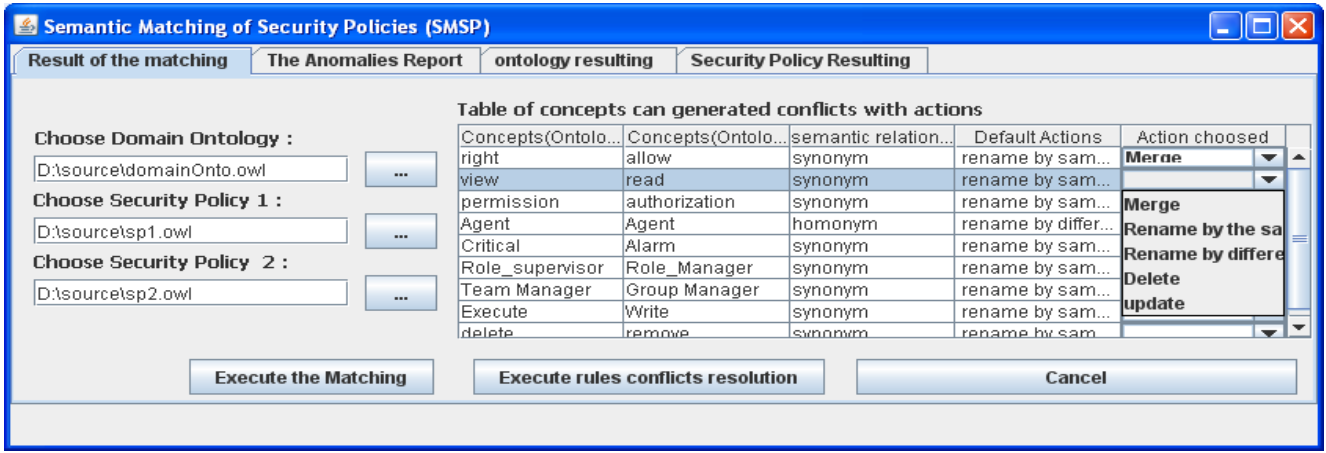

Figure 5. A prototype of Semantic Matching of Security Policies

## VII. CONCLUSION AND FUTURE WORK

We were interested in this work to conflict resolution semantic naming type when aligning SP in distributed multi-systems security policies. Our solution is based on ontologies mediation for SP cooperation and understanding. We propose a helpful process for security experts, this process permit to mask heterogeneity and resolve conflicts between SP, using different steps. The first and last steps concern the processing of conceptual ontological policies representations. The second step, which constitutes the fundamental part of this work, is a detecting method for the semantic conflicts between SP. We expect to continue this work first by a formal validation of the solution, and then by looking for opportunities to extend it to solve other types of semantic conflicts, including conflicts of measurement and confusion. We plan to further our research on the use of this approach in CC environments toward a centralised management of security policies, we will propose a framework that can interface with other security-based frameworks to ensure the interoperability and also enrich them with semantic relationships.